\documentclass[twocolumn,amsmath,amssymb,nofootinbib,floatfix]{revtex4}

\usepackage[pagewise]{lineno}
\usepackage{graphicx}
\usepackage{dcolumn}
\usepackage{bm}
\usepackage{hyperref}
\usepackage{comment}
\usepackage{color}
\usepackage{subfigure}
\usepackage[papersize={10.5in,12in}]{geometry}
\usepackage{appendix}

\makeatletter

\newcommand{\Rmnum}[1]{\expandafter\@slowromancap\romannumeral #1@}
\makeatother

\def\be{\begin{equation}}
\def\ee{\end{equation}}
\def\bea{\begin{eqnarray}}
\def\eea{\end{eqnarray}}
\def\bal{\begin{aligned}}
\def\eal{\end{aligned}}

\begin{document}
\title{Exploring Quantum Aspects of Dark Matter Axions and Dark Photons Transitioning to Photons in a Resonant Cavity}
\author{Ruifeng Zheng}
\author{Puxian Wei}
\author{Qiaoli Yang}
\email{qiaoliyang@jnu.edu.cn}

\affiliation{Department of Physics, College of Physics and Optoelectronic Engineering, Jinan University, Guangzhou 510632, China}	

\begin{abstract}
When axion cold dark matter interacts with a static magnetic field, it can be converted to photons with energy near the axion's mass. Classical analysis shows that incorporating a resonant cavity significantly enhances this conversion rate, forming the basis for many experiments aimed at detecting dark matter axions. However, one might ask: Does the axion-photon conversion rate increase for a single axion-photon transition? Answering this question could lead to optimizing the search for axions by integrating quantum measurement techniques. In this paper, we demonstrate that at the quantum level, single axion-photon transitions are amplified by the cavity quality factor $Q$. Furthermore, the coherence of dark matter waves is unnecessary during the measurement. The underlying principle is similar to the Purcell effect. Additionally, we provide an analysis of the scenario involving dark photon dark matter.
\end{abstract}

\maketitle
\section{Introduction}
\label{sec1}
Cosmological observations suggest that dark matter makes up around $27\%$ of the total energy density of the universe. Despite its abundance, the exact nature of dark matter remains a mystery; current understanding indicates that it cannot be composed of the Standard Model particles. This puzzle has sparked extensive investigations into various dark matter candidates. Among these, axions and dark photons have emerged as prominent bosonic candidates \cite{Peccei:1977hh, Peccei:1977ur,Wilczek:1977pj,Weinberg:1977ma,Kim:1979if, Shifman:1979if, Zhitnitsky:1980tq, Dine:1981rt, Abbott:1982af, Dine:1982ah, Sikivie:1982qv,Preskill:1982cy,Wilczek:1991jgb,Berezhiani:1992rk,Covi:2001nw,Sikivie:2006ni,Hertzberg:2008wr, Duffy:2009ig, Nelson:2011sf,Collar:2012olx,Marsh:2015xka,Klaer:2017ond, Agrawal:2018vin, Caputo:2021eaa}, attracting considerable interest within the scientific community.

Axions arise as pseudo-Nambu-Goldstone bosons following the spontaneous and explicit breaking of the $U(1)$ Peccei-Quinn (PQ) symmetry. They acquire a small mass from the Quantum Chromodynamics (QCD) sector by the instanton effect \cite{Callan:1977gz,Crewther:1979pi,Vafa:1984xg}. Their exceptionally weak interaction with the Standard Model particles makes them highly elusive to detect. Dark photons, on the other hand, are massive gauge bosons associated with a newly introduced $U(1)$ gauge symmetry \cite{Holdom:1985ag} in hidden sectors. Despite their nearly collision-less behavior, dark photons exhibit a slight mixing with visible photons. The formation of axion or dark photon dark matter can be attributed to the intriguing "misalignment mechanism" in the early universe. This mechanism resulted in axion or dark photon dark matter accounting for a significant portion of the observed dark matter abundance today. Currently, there are many proposed or ongoing detection experiments actively searching for dark matter axions and dark photons \cite{DePanfilis:1987dk, Hagmann:1990tj, Kahn:2016aff, Marsh:2018dlj, Schutte-Engel:2021bqm,Kim:2023vpo,Quiskamp:2023ehr,Yang:2023yry,APEX:2024jxw,DiVora:2024ikr} etc.

Classical analysis has shown that the conversion rate between axions and photons within a static magnetic field can be substantially enhanced by using a resonant cavity \cite{Sikivie:1983ip,Sikivie:1985yu,Krauss:1985ub}. The conversion power is determined by the volume of the cavity, its quality factor $Q_c$, and the strength of the static magnetic field. This theoretical framework can be extended to the case of dark photon to photon transitions. The underlying principle in these analyses is the resonance between axion or dark photon fields with a photon field within the cavity. One notable difference between the two processes is that axions interact with two photons, while dark photons only interact with one. As such, axion-photon conversion necessitates an external magnetic field, whereas dark photon to photon transitions do not require it, simplifying experimental requirements for the latter. For instance, the ADMX experiment \cite{ADMX:2021nhd, ADMX:2020ote}, designed to search for axion dark matter, can also be repurposed to detect dark photons. Recently, several studies have suggested the use of cavity haloscopes to look for both types of transition \cite{Caldwell:2016dcw, HAYSTAC:2018rwy, He:2024ytp, BREAD:2023xhc, He:2024fzj}.

However, the classical picture may have limitations. Due to the exceedingly weak interaction of axions or dark photons with photon fields, the power converted in a typical resonant cavity is equivalent to that of a solitary photon per second. It is not clear whether the boost occurs at the single-photon level, which could be crucial for future experiments with quantum sensors. Certainly, the distinction between 'quantum' and 'classical' here is somewhat academic, as leading-order effects of quantum field perturbation theory can also be regarded as a 'classical' effect.

In this paper, we show that the single axion or dark photon transition rate indeed increases due to the cavity's quality factor $Q_c$, attributable to the final photon wave function being confined within the cavity, much like the Purcell effect. This could facilitate the integration of quantum measurement techniques into future experiments \cite{Yang:2022uil}. Moreover, the transition rate is independent of the coherence of dark matter particles, similar to multiply emitters in the Purcell effect, thus validating the cavity method for a broader range of scenarios (for example, searching for narrow-bandwidth noncoherent dark matter, of which spatial variation remains little in the cavity scale, but the wave amplitude oscillates randomly: $D=D_0e^{i[\omega_0 t +\phi(t)]}$, with $\phi(t)$ a random function). The quantum transition rate is $\pi/2$ higher than the classical scenario in our calculation. This could be understood because terms such as $\pi/2$ generally appear in quantum transitions, such as in Fermi's golden rule. It results from the delta function that ensures the energy conservation between two quantum states, while in the classical case the environmental damping transfers the delta function to a Lorentzian distribution that assimilates $\pi/2$. When a cavity is populated with many photons, the photon-coherent state in the cavity evolves as: $|\phi(t)\rangle=|\alpha e^{-i\omega t- \kappa t/2}\rangle$, indicating that the amplitude of the coherent states damps at a rate of $\kappa/2$. Then the delta function becomes a Lorentzian and the result merges to the classical result.       

\section{Quantum Perspectives of Axion-Photon transitions inside a Resonant Cavity}
\label{sec234}
Dark matter halos are one of the fundamental building blocks that make up cosmic structures. The halos surround galactic disks and extend far beyond the edges of visible galaxies, forming a region primarily composed of dark matter. Although no direct observations of dark matter halos have been made thus far, their size could provide valuable insights into the properties of dark matter particles. For example, the typical de Broglie wavelength of dark matter should not exceed the scale of galaxies, which sets a boundary for the mass ($\sim10^{-22}$~eV) of dark matter particles. Indeed, the phase-space distribution of dark matter halos is crucial to both cosmological structure formation and dark matter searches.

The spatial distribution of dark matter is relatively well-established. According to Planck 2018 observations \cite{Planck:2018vyg}, the average abundance of cold dark matter in the universe is $\Omega_ch^2=0.1200\pm0.0012$ where $h=H_0/100~\mathrm{km}\cdot \mathrm{s} ^{-1}\cdot \mathrm{Mpc} ^{-1}$, and $H_0$ is the Hubble constant. The energy density of the local dark matter halo of our galaxy is approximately $\rho_{DM}\approx0.43{\rm GeV/cm^3}$ \cite{Nesti:20133P}. If dark matter consists of axions or dark photons, its particle number density will be extremely high. For example, for axions with a mass of $\mu$eV, the number density would reach $n=\rho_{DM}/m_a\approx 10^{14}/{\rm cm^3}$.

The velocity distribution of dark matter halos indeed has many more uncertainties; e.g., some models result in a more coherent dark matter wave distribution such as axions BEC \cite{Sikivie:2009qn}, For the purposes of this paper, it suffices to show that coherence is not required for enhanced transitions. Consequently, we adopt a stochastic wave function model \cite{Foster:2017hbq}:
\begin{equation}\label{EQ1}
    \begin{aligned}
D(t)=&\frac{\sqrt{\rho_{DM} }}{m} \sum_{j}^{} \alpha _j\sqrt{f(v_j)\Delta v} \\&\times \cos \left [m\left ( 1+\frac{v_{j}^{2} }{2}  \right )t+\phi _{j}    \right ] ~,
    \end{aligned}
\end{equation}
where $\rho_{DM}\approx0.43$ GeV/cm$^3$ is the local energy density of dark matter, $\alpha _j$ is the random number of Rayleigh distribution, $\phi_{j}$ is a random phase, and $f(v_j)$ is the velocity distribution of dark matter. In general cases, the wave could be completely incoherent.

The interaction Lagrangian between axions and photons is given by:
\begin{equation}\label{eq11}
\mathcal{L}_{a\gamma \gamma}=g_{a\gamma \gamma} a\vec{\mathbf{E}} \cdot \vec{\mathbf{B}}_0~,
\end{equation}
where $a$ is the local axion field, and $g_{a\gamma\gamma}$ represents the axion-photon coupling constant. This constant can be expressed as: $g_{a\gamma\gamma}=\alpha c_{\gamma }/(2\pi f_{a})$ \cite{2021JHEP...01..172C}. Here, $f_a$ is the axion decay constant (approximately $10^{12}\sim 10^{16}$ GeV), $\alpha=1/137$ denotes the fine-structure constant, and $c_{\gamma}$ represents a model-dependent constant of order one. Moreover, $\vec {\bf E}$ is the electric field of photons inside the cavity, while $\vec {\bf B}_0$ corresponds to an applied static external magnetic field.

Utilizing Eq.(\ref{EQ1}), we can express the local axion field as:
\bea\label{Eq99}
a(t)&=&\frac{\sqrt{\rho_{\rm DM}} }{m_a} \sum_{j}^{} \alpha _j\sqrt{f(v_j)\Delta v}\nonumber\\ &\times& \cos \left[m_a\left(1+\frac{v_{j}^{2} }{2}\right)t+\phi _{j}\right] ~.
\eea
For simplicity, we can assume that the applied external magnetic field is aligned in the $z$ direction: $\vec {\bf B}_0=B_0\hat z$. Modern cryogenic technology can maintain a temperature of several mK or lower inside the cavity. This leads to a subunity thermal photon occupation number at the resonant frequency, typically around several GHz. As such, it is crucial to consider the quantum properties of the photon field inside the cavity. The electric field $\vec{\mathbf{E}}$ of a photon inside a cavity can be expressed in terms of creation operators $a_k^\dagger$ and annihilation operators $a_k$, with cavity mode wave functions $\vec{\textbf u}_{k}(\vec x)$ as:
\bea\label{caeq}
\vec{\bf E} =i\sum_k\sqrt{\frac{\omega_k} {2}}(a_k\vec{\bf u}_ke^{-i\omega_kt}-a_k^{\dagger}\vec{\bf u}_k^*e^{i\omega_kt})~,
\eea
where $\omega_k$ denotes the wave frequency of the cavity mode $k$ and $\vec{\textbf u}_{k}$ satisfies $(\nabla^2+\omega_k^2)\vec{\textbf u}_{k}(\vec x)=0$ as well as appropriate boundary conditions of the cavity.

The Hamiltonian of the axion-photon interaction is given by
\bea
H_{I}&=&-\int d^{3}x\mathcal{L}_{a\gamma \gamma }=-g_{a\gamma \gamma}B_0a(t) \int d^3x \vec{\mathbf{E} }\cdot\hat{z}\nonumber\\
&=&-\frac{ig_{a\gamma \gamma}B_0\sqrt{\rho_{DM}} }{\sqrt{2} m_a}\cdot \sum_{j,k} \alpha _j\sqrt{f(v_j)\Delta v} \nonumber\\ &\times& \cos (\omega_jt+\phi _{j} )
\sqrt{\omega _k} (a_ke^{-i\omega _kt}U_k-c.c.)~.
\eea
Here, $U_{k}=\int d^3x\vec{\bf u}_k \cdot \hat{z}$, $\omega_{j}=m_a\left ( 1+v_{j}^{2}/2  \right )$, and $c.c.$ denotes the conjugate of the complex. The leading order transition probability $P$ for cavity photon state $ \left | 0 \right \rangle \to  \left | 1 \right \rangle $ is expressed as
\bea\label{EQ21}
P&\approx& \left |\left \langle 1\left | \int_{0}^{t}  dtH_{I} \right | 0 \right \rangle   \right | ^{2}
=\frac{g_{a\gamma \gamma}^2B_0^2\rho_{DM} }{2m_a^2}\nonumber\\ &\times&(\sum_{j,k}\sqrt{\omega_{k}} \alpha _j \sqrt{f(v_j)\Delta v} U_{k}^{*}    \int_{0}^{t}dt\cos \left (\omega_{j}t+\phi _{j}    \right ) e^{i\omega_{k}t} )\nonumber\\&\times& c.c.~~,
\eea
where we have assumed that $v_j\sim~$constant during a single axion-photon transition. The terms with poles such as $\omega_j=\omega_k$ are dominate and the $j\neq j'$, $k\neq k'$ terms vanish, leading to
\begin{equation}\label{EQ2111}
\begin{aligned}
&\sum_{jj'kk'}\int_{0}^{t}dt\cos(\omega_{j}t+\phi _{j} ) e^{i\omega_{k}t}\int_{0}^{t}dt\cos (\omega_{j'}t+\phi _{j'}  ) e^{-i\omega_{k'}t} \\
\approx&\sum_{jj'kk'}\frac{1}{4} e^{-i(\phi_j-\phi_{j'})}\frac{ [ e^{i(\omega _k- \omega_j)t}-1 ]  [ e^{-i\left(\omega _{k'}- \omega_{j'}\right)t}-1  ]   }{(\omega _k-\omega _j  ) ( \omega _{k'}-\omega _{j'}  )  } \\
\approx&\sum_{j,k}\frac{\sin ^2[(\omega _k-\omega _j)t/2]}{4[(\omega _k-\omega _j)/2]^2} ~.
 \end{aligned}
\end{equation}
When resonance occurs, \( \omega_k - \omega_j < \delta\omega = \omega_0/Q \sim 10^5 \, \text{Hz} \) for most axion cavities, corresponding to a timescale of \( 10^{-5} \, \text{s} \). In contrast, the photon-axion transition rate is typically 1 event per second, which corresponds to a time scale of \( t \sim 1 \, \text{s} \). Thus $t$ is much larger than $|1/(\omega_k-\omega_j)|$, we can use the approximation ${\rm sin}^2(\Delta\omega t/2)/(\Delta \omega/2)^2\approx 2\pi t\delta(\Delta \omega )$. We also define the form factor as:
\begin{equation}
C_k =\frac{\left |\int d^3x \vec {\bf u}_k \cdot \hat{z}   \right | ^2}{V\int d^3x\left | \vec {\bf u}_k   \right |^2 } ~,
\end{equation}
where $\int d^3x|\vec {\bf u}_k|^2$ is used to properly normalize the electric field operators. The quality factor of a cavity $Q_c=\omega_0/\Delta \omega$ where $\omega_0$ is the resonance frequency and $\Delta \omega$ is the frequency bandwidth of the cavity. The quality factor of dark matter axions $Q_a\approx1/\delta v_a^2$. Combining these factors, we can express the transition probability of axion-photon in a cavity as
\begin{equation}\label{EQ111777}
\begin{aligned}
P&=\frac{\pi g_{a\gamma \gamma}^2B _0^2\rho_{DM} tV}{4m_a^2}\sum_{jk}C_k\delta (\omega_k-\omega _j)\omega_k \alpha _j^2 f(v_j)\Delta v  \\
\approx&\frac{\pi g_{a\gamma \gamma}^2B_0^2\rho_{DM} tV\langle\alpha ^2\rangle}{4m_a^2}\int dvf(v)\int d\omega_k   C_k\frac{\omega_k}{d\omega_k} \delta (\omega_k-\omega_{v})\\
\approx&\frac{\pi g_{a\gamma \gamma}^2B_0^2tV}{2m_a^2}\rho_{DM} \int_0^{\Delta} dv_a f(v_a)C_{\omega_a}Q_c~,
\end{aligned}
\end{equation}
where $\left \langle \alpha ^2 \right \rangle =\int \alpha^2 P(\alpha)d\alpha=2$ represents the expectation value for $\alpha$. $\Delta\approx \sqrt{2/Q_c}$ is the velocity of axions within the cavity's bandwidth. Assume $f(v)$ is sharply peaked near the cavity resonance $\int f(v)dv \approx 1$. The transition rate is then
\begin{equation}
\begin{aligned}
R=\frac{dP}{dt} \approx \frac{\pi g_{a\gamma \gamma}^2B_0^2\rho_{DM} V}{2m_a^2} Q_cC_{\omega_a}~.
\end{aligned}
\end{equation}
Consequently the axion-photon transition power in a cavity is
\begin{equation}\label{EQ211}
\begin{aligned}
Pw=\omega_a R\approx \frac{\pi g_{a\gamma \gamma}^2B_0^2\rho_{DM} V}{2m_a}Q_cC_{\omega_a}~.
\end{aligned}
\end{equation}
Eq.(\ref{EQ211}) indicates that the power of a single-axion-photon transition is indeed enhanced by the presence of a resonant cavity. When there is an ambient bath of thermal photons with a photon occupation number $n$ in the cavity, it may seem tempting to think that the conversion rate could be higher because the raising operator $ a^\dagger $ acting on the state $|n\rangle$ produces $ \sqrt{n+1} |n+1\rangle $. However, the reverse process of photons to axions cancels out the boosting effect of higher photon occupation numbers. 

\section{Quantum Perspective on Dark Photon to Photon Transitions in a Resonant Cavity}\label{sec4}
Dark matter dark photons are hidden U(1) vector bosons with a small mass and very weak mixing to the Standard Model photons. Their mass could be a Stueckelberg mass, which naturally arises for an Abelian gauge boson in low-energy effective field theory. Additionally, dark photons can be non-thermally created in the early universe, potentially comprising a significant fraction of dark matter today. The general Lagrangian describing two U(1) bosons and their mixing is given by:
\bea
\begin{aligned}
\mathcal{L}=&-\frac{1}{4} F^{\mu\nu} F_{\mu\nu}-\frac{1}{4} F'^{\mu\nu} F'_{\mu\nu}-\frac{1}{2}\chi F^{\mu\nu} F'_{\mu\nu} \\
&+\frac{1}{2} m_{\gamma^{\prime}}^{2} A'_{\mu} A'^{\mu}+J^{\mu}A_{\mu}~.
\end{aligned}
\eea
Here, $F_{\mu\nu}$ represents the field strength tensor for the ordinary U(1) field $A_{\mu}$, while $J^{\mu}$ denotes the standard electric current. The hidden U(1) field $A'_{\mu}$ has a corresponding field strength tensor $F'_{\mu\nu}$. The third term corresponds to a non-diagonal kinetic mixing term, which generally exists after integrating out UV quantum processes of a more fundamental theory. In compactifications of string theory, $\chi$ in the $10^{-12}\sim10^{-3}$ range have been predicted in the literature \cite{Dienes:1996zr,Goodsell:2009xc}.

To understand the properties of dark matter dark photons, we can switch to the propagating basis or mass eigenstates in which the photon and dark photon decouple in vacuum. This is achieved using the following transformation: $A_{\mu} \rightarrow A_{\mu} - \chi A'_{\mu}$, and $A'_{\mu} \rightarrow A'_{\mu}+{\cal O}(\chi^2)$. The Lagrangian then becomes:
\bea
\bal
{\mathcal{L}}=&-\frac{1}{4}{F}_{\mu\nu}{F}^{\mu\nu}-\frac{1}{4}{F'}_{\mu\nu}{F'}^{\mu\nu}\\&+\frac{m_{\gamma^{\prime}}^{2}}{2}A'_{\mu}A'^{\mu}+J^{\mu}\left({A}_{\mu}-\chi A'_{\mu}\right).
\eal
\eea
After creation, dark matter dark photons can be regarded as free-streaming in the universe. Using the Lorenz gauge condition $\partial^{\mu}A'_{\mu}=0$, the dark field obeys the wave equation: $(\partial^2 + m_{\gamma'}^2) A'_{\mu} = 0$. The time-like polarization $A'_0$ is suppressed by the fixing condition, so it can be neglected. Thus, dark matter photons are dominated by $\vec A'$. Furthermore, $\vec{E'} = -\partial \vec{A'}/\partial t \approx -im_{\gamma'}\vec{A'}$, and $B'\sim v E'\ll E'$. Therefore, the dark photons are mainly composed of their electric field component.

When a resonant cavity is present, the dark electric field can induce a weak electric current that generates photons inside the cavity. To calculate this effect, we can rotate the vector fields to the flavor eigenstates by: \(A_{\mu} \rightarrow A_{\mu} - \chi A'_{\mu}\) and
\(A'_{\mu} \rightarrow A'_{\mu} + \chi A_{\mu}\). This yields a Lagrangian with a dark photon-photon interaction term, while the dark photons do not couple to electric current up to $\chi^2$:
\bea
\bal
{\mathcal{L}}=&-\frac{1}{4}{F}_{\mu\nu}{F}^{\mu\nu}-\frac{1}{4}{F'}_{\mu\nu}{F'}^{\mu\nu}\\&+\frac{m_{\gamma^{\prime}}^{2}}{2}A'_{\mu}A'^{\mu}+\chi m_{\gamma'}^2A_{\mu}A'^{\mu}+J^{\mu}{A}_{\mu}.
\eal
\eea
These two pictures are equivalent, and the latter is more convenient for subsequent calculations. The dark photon-photon interaction term is then:
\be
{\cal L}_I = -\chi m_{\gamma'}^2 {A}_{\mu} {A}'^{\mu}
\ee
where $\chi$ denotes the kinetic mixing parameter and $m_{\gamma'}$ is the mass of the dark photon.

As $A'_0 \approx 0$, the dark photon-photon interaction can be approximated as:
\[{\cal L}_I \approx -\chi\vec{E} \cdot \vec{E'}\]
The polarization of the dark photon can be aligned or point in random directions depending on its creation mechanism and later structure formation. The difference may be not significant, but for the latter scenario, one needs to add a small wave amplitude factor. Assuming the first case, where $\vec{A'} \propto \hat{n} {\sqrt{2\rho_{DM}}/m_{\gamma'}}$, with $\hat{n}$ being a unit vector with fixed direction, Eq.~(\ref{EQ1}) yields:
\begin{equation}
 \begin{aligned}
\vec{E}^{\prime } =&-i\frac{\partial \vec A' }{\partial t} \\ \approx&-i\sqrt{\rho_{DM}}\sum_{j}^{} \alpha _j\sqrt{f(v_j)\Delta v}\\& \times \cos \left [m_{\gamma'}\left ( 1+\frac{v_{j}^{2} }{2}  \right )t+\phi _{j}    \right ]\hat{n}  ~,
\end{aligned}
\end{equation}
Similar to the previous section, let us define $\omega_j = m_{\gamma'} (1 + v^2_j / 2)$. The electric field inside a cavity $\vec{\bf E}$ can be expanded into creation and annihilation operators. The Hamiltonian of the interaction is then:
 \begin{equation}
 \begin{aligned}
H_I=&-\int d^3x\mathcal{L} _I\\
=&\chi \sqrt{\rho_{DM} }\cdot \sum_{jk}^{} \alpha _j\sqrt{f(v_j)\Delta v} \cos \left (\omega_jt+\phi _{j}    \right )\\
&\times \sqrt{\omega _k  2}\left ( a_ke^{-i\omega _kt}U_{k}-c.c.\right ) ~,
\end{aligned}
\end{equation}
where $U_{k} = \int d^3x\vec{\bf u}_k \cdot \hat{n}$. The transition probability of a cavity photon from the ground state $|0\rangle$ to an excited state $|1\rangle$ can be expressed as:
 \begin{equation}\label{CCDPp}
 \begin{aligned}
P\approx& \left |\left \langle 1\left | \int_{0}^{t}  dtH_{I} \right | 0 \right \rangle   \right | ^{2}=\frac{\pi\chi ^2\rho_{DM} tV}{4}\int \alpha _v ^2 f(v)d v \\\times& \int \mathbf{C} _k\frac{\omega_k}{d\omega} d\omega \delta \left (\omega_k-\omega_{\gamma'}\right)\approx\frac{\pi\chi ^2\rho_{DM} tV\mathbf{C} _\omega}{2}\\\times& Q_c~,
\end{aligned}
\end{equation}
where the form factor is defined as:
\begin{equation}
 \begin{aligned}
\mathbf{C} _k=\frac{\left | \int d^3x\vec{\textbf u}_{k}\cdot \vec{n} \right |^2}{V\int d^3x\left | \vec{\textbf u}_{k} \right |^2 }~,
\end{aligned}
\end{equation}
and we have assumed that the dark photon velocity distribution is sharply peaked near the cavity resonance.
The transition rate is given by:
\begin{equation}
 \begin{aligned}
R=\frac{dP}{dt}=\frac{\pi \chi ^2\rho_{DM} V\mathbf{C}_{\omega}}{2}Q_c~,
\end{aligned}
\end{equation}
and the dark photon-induced cavity power is:
 \begin{equation}\label{EQ311}
 \begin{aligned}
Pw=\omega_DR\approx\frac{\pi \chi ^2m_{\gamma'}\rho_{DM} V\mathbf{C}_{\omega}}{2}Q_c~.
\end{aligned}
\end{equation}

\section{conclusions}\label{sec7}
Light bosonic particles can be produced in the early universe via nonthermal production mechanisms such as vacuum misalignment. If their interactions with Standard Model particles are weak, their relics can constitute cold dark matter. A large region of their parameter space can give rise to a substantial fraction of the observed dark matter density, which presents both opportunities and challenges to dark matter searches.

The era of quantum sensing can mitigate the challenges by leveraging sensitivity to unprecedented single-particle levels. The cavity-enhanced experiments are particularly attractive in this regard because the single axion or dark photon to photon transition rate is enhanced by confining the final photon states inside a cavity. This enhancement remains even if the dark matter particles are not in a coherent state during measurement. The insight is that the transition rate enhancement arises from the confinement of the photon in the cavity (Purcell effect), which amplifies the density of photon states near resonance. This enhancement depends on the cavity quality factor $Q_c$ and is independent of dark matter phase coherence. Even if individual axions or dark photons lack phase coherence, their contributions to the transition rate add constructively. This is analogous to how spontaneous emission of radiative materials into a cavity is enhanced regardless of the coherence of the emitters. The application scenarios for cavity experiments therefore could be broader, and additional search strategies, especially incorporating quantum sensing technologies, are encouraging.

Future studies of the dark matter halo structures will also be crucial. If axions or dark photons dark matter exhibit extremely narrow bandwidths or a high inherent quality factor $Q_a\gg Q_c$, such as they are in a Bose-Einstein condensate, the experiments can benefit from using high Q cavities, potentially including state-of-the-art superconducting types. Certainly, the bandwidth covered at each frequency pin is narrowed, but the signal improves dramatically, which more than compensates for the loss of bandwidth.

\section*{ACKNOWLEDGMENTS}
We thank Yu Gao, Nick Houston, Danning Li, and Zhihui Peng for valuable discussions. This work has been supported in part by the NSFC under Grant No.12150010.
\bibliography{dp}
 
\end{document}